\documentclass[12pt,a4paper]{article}
\usepackage[cp850]{inputenc}
\usepackage{amssymb}
\usepackage{latexsym}
\usepackage[dvips]{epsfig}
\usepackage{pstricks,pst-node}
\renewcommand{\baselinestretch}{1.5}
\begin{document}
\begin{titlepage} 
\renewcommand{\baselinestretch}{1}
\small\normalsize
\begin{flushright}
hep-th/0511260\\
MZ-TH/05-26     \\
\end{flushright}

\vspace{0.6cm}

\begin{center}   

{\LARGE \sc Asymptotic Safety in \\[1mm] Quantum Einstein Gravity:\\[5mm] 
nonperturbative renormalizability and\\[3.5mm] fractal spacetime 
structure\footnote{Invited paper at the Blaubeuren Workshop 2005 on
Mathematical and Physical Aspects of Quantum Gravity.}}

\vspace{2cm}
{\large O. Lauscher$^{(a)}$ and M. Reuter$^{(b)}$}\\

\vspace{1.7cm}
\noindent
{(a)\it Institute of Theoretical Physics, University of Leipzig\\
Augustusplatz 10-11, D-04109 Leipzig, Germany}\\[12pt]
{(b)\it Institute of Physics, University of Mainz\\
Staudingerweg 7, D-55099 Mainz, Germany}\\[6pt]
\end{center}   
 
\vspace*{1.6cm}

\begin{abstract}
The asymptotic safety scenario of Quantum Einstein Gravity, the quantum field
theory of the spacetime metric, is reviewed and it is argued that the theory is
likely to be nonperturbatively renormalizable. It is also shown that asymptotic
safety implies that spacetime is a fractal in general, with a fractal dimension
of 2 on sub-Planckian length scales.
\end{abstract}
\end{titlepage}
\section{Introduction}
\renewcommand{\theequation}{1.\arabic{equation}}
\setcounter{equation}{0}
\label{3intro}
Quantized General Relativity, based upon the Einstein-Hilbert action 
\begin{eqnarray}
\label{l1}
S_{\rm EH}=\frac{1}{16\pi G}\int d^4x\,\sqrt{-g}\,\left\{-R+2\Lambda
\right\}\;,
\end{eqnarray}
is well known to be 
perturbatively nonrenormalizable. This has led to the widespread believe that
a straightforward quantization of the metric degrees of freedom cannot lead to
a mathematically consistent and predictive {\it fundamental} theory valid down
to arbitrarily small spacetime distances. Einstein gravity was rather
considered merely an {\it effective} theory whose range of applicability is
limited to a phenomenological description of gravitational effects at
distances much larger than the Planck length.

In particle physics one usually considers a theory fundamental if it is
perturbatively renormalizable. The virtue of such models is that one can
``hide'' their infinities in only finitely many basic parameters (masses,
gauge couplings, etc.) which are intrinsically undetermined within the theory
and whose value must be taken from the experiment. All other
couplings are then well-defined computable functions of those few parameters.
In nonrenormalizable effective theories, on the other hand, the divergence
structure is such that increasing orders of the loop expansion require an
increasing number of new counter terms and, as a consequence, of undetermined
free parameters. Typically, at high energies, all these unknown parameters
enter on an equal footing so that the theory looses its predictive power.

However, there are examples of field theories which do ``exist'' as fundamental
theories despite their perturbative nonrenormalizability \cite{wilson,parisi}.
These models are ``nonperturbatively renormalizable'' along the lines of
Wilson's modern formulation of renormalization theory \cite{wilson}. They are
constructed by performing the limit of infinite ultraviolet cutoff
(``continuum limit'') at a non-Gaussian renormalization group fixed point
${\rm g}_{*i}$ in the space $\{{\rm g}_i\}$ of all (dimensionless, essential) 
couplings ${\rm g}_i$ which parametrize a general action functional. This 
construction has 
to be contrasted with the standard perturbative renormalization which, at least
implicitly, is based upon the Gaussian fixed point at which all couplings
vanish, ${\rm g}_{*i}=0$ \cite{polch,bagber}.

\section{Asymptotic safety}
\renewcommand{\theequation}{2.\arabic{equation}}
\setcounter{equation}{0}
\label{as}
In his ``asymptotic safety'' scenario Weinberg \cite{wein} has put
forward the idea that, perhaps, a quantum field theory of gravity can be
constructed nonperturbatively by invoking a non-Gaussian ultraviolet (UV)
fixed point $({\rm g}_{*i}\neq 0)$. The resulting theory would be 
``asymptotically
safe'' in the sense that at high energies unphysical singularities are likely
to be absent. 

The arena in which the idea is formulated is the so-called ``theory space''.
By definition, it is the space of all action functionals $A[\,\cdot\,]$ 
which
depend on a given set of fields and are invariant under certain symmetries. 
Hence the theory space $\{A[\,\cdot\,]\}$ is fixed once the field contents and 
the symmetries are fixed. The infinitely many generalized couplings 
${\rm g}_i$ needed to
parametrize a general action functional are local coordinates on theory space.
In gravity one deals with functionals $A[g_{\mu\nu},\cdots]$ which are required
to depend on the metric in a diffeomorphism invariant way. (The dots represent
matter fields and possibly background fields introduced for technical 
convenience.) Theory space carries
a crucial geometric structure, namely a vector field which encodes the
effect of a Kadanoff-Wilson-type block spin or ``coarse graining'' procedure,
suitably reformulated in the continuum. The components {\boldmath$\beta$}{$_i$}
of this vector field are the beta-functions of the couplings ${\rm g}_i$. They
describe the dependence of ${\rm g}_i\equiv{\rm g}_i(k)$ on the coarse 
graining scale $k$:
\begin{eqnarray}
\label{rgrho}
k\,\partial_k{\rm g}_i=\mbox{\boldmath$\beta$}_i({\rm g}_1,{\rm g}_2,\cdots)
\end{eqnarray}
By definition, $k$ is taken to be a mass scale. Roughly speaking the running 
couplings
${\rm g}_i(k)$ describe the dynamics of field averages, the averaging volume
having a linear extension of the order $1/k$. The ${\rm g}_i(k)$'s should be
thought of as parametrizing a running action functional $\Gamma_k[g_{\mu\nu},
\cdots]$. By definition, the renormalization group (RG) trajectories, i.e. the
solutions to the ``exact renormalization group equation'' (\ref{rgrho}) are
the integral curves of the vector field $\vec{\mbox{\boldmath$\beta$}}
\equiv(\mbox{\boldmath$\beta$}_i)$ defining the ``RG flow''.

The asymptotic safety scenario assumes that $\vec{\mbox{\boldmath$\beta$}}$
has a zero at a point with coordinates ${\rm g}_{*i}$ not all of which are
zero. Given such a non-Gaussian fixed point (NGFP) of the RG flow one defines 
its UV critical surface ${\mathcal S}_{\rm UV}$ to consist of all points of 
theory space which are
attracted into it in the limit $k\rightarrow\infty$. (Note that
increasing $k$ amounts to going in the direction {\it opposite} to the natural
coarse graining flow.) The
dimensionality ${\rm dim}\left({\mathcal S}_{\rm UV}\right)\equiv 
\Delta_{\rm UV}$
is given by the number of attractive (for {\it in}creasing cutoff $k$) 
directions in the space of couplings. The linearized flow near the fixed point
is governed by the Jacobi matrix
${\bf B}=(B_{ij})$, $B_{ij}\equiv\partial_j
\mbox{\boldmath$\beta$}_i({\rm g}_*)$:
\begin{eqnarray}
\label{p2.1}
k\,\partial_k\,{\rm g}_i(k)=\sum\limits_j B_{ij}\,\left({\rm g}_j(k)
-{\rm g}_{*j}\right)\;.
\end{eqnarray}
The general solution to this equation reads
\begin{eqnarray}
\label{p2.8}
{\rm g}_i(k)={\rm g}_{*i}+\sum\limits_I C_I\,V^I_i\,
\left(\frac{k_0}{k}\right)^{\theta_I}
\end{eqnarray}
where the $V^I$'s are the right-eigenvectors of ${\bf B}$ with (complex) 
eigenvalues $-\theta_I$. Furthermore, $k_0$ 
is a fixed reference scale, and the $C_I$'s are constants of integration. If
${\rm g}_i(k)$ is to approach ${\rm g}_{*i}$ in the infinite cutoff limit
$k\rightarrow\infty$ we must set $C_I=0$ for all $I$ with 
${\rm Re}\,\theta_I<0$. Hence the dimensionality $\Delta_{\rm UV}$ equals the 
number of ${\bf B}$-eigenvalues with a negative real part, i.e. the number of
$\theta_I$'s with a positive real part.

A specific quantum field theory is defined by a RG trajectory which exists
globally, i.e. is well behaved all the way down from ``$k=\infty$'' in the UV
to $k=0$ in the IR. The key idea of asymptotic safety is to base the theory 
upon one of the trajectories running inside the hypersurface 
${\mathcal S}_{\rm UV}$ since these trajectories are manifestly well-behaved
and free from fatal singularities (blowing up couplings, etc.) in the large$-k$
limit. Moreover, a theory based upon a trajectory inside 
${\mathcal S}_{\rm UV}$ can be predictive, the problem of an increasing number
of counter terms and undetermined parameters which plagues effective theory
does not arise. 

In fact, in order to select a specific quantum theory
we have to fix $\Delta_{\rm UV}$ free parameters which are not predicted
by the theory and must be taken from experiment. When
we {\it lower} the cutoff, only $\Delta_{\rm UV}$ parameters
in the initial action are ``relevant'', and fixing these parameters
amounts to picking a specific trajectory on
${\mathcal S}_{\rm UV}$; the remaining ``irrelevant'' parameters 
are all attracted towards ${\mathcal S}_{\rm UV}$ automatically. Therefore the 
theory has the more predictive power the smaller is the dimensionality of
${\mathcal S}_{\rm UV}$, i.e. the fewer UV attractive eigendirections the
non-Gaussian fixed point has. If $\Delta_{\rm UV}<\infty$, the quantum field 
theory thus constructed is as predictive as a perturbatively
renormalizable model with $\Delta_{\rm UV}$ ``renormalizable couplings'', i.e.
couplings relevant at the Gaussian fixed point.

It is plausible that ${\mathcal S}_{\rm UV}$ is indeed finite dimensional. If 
the dimensionless ${\rm g}_i$'s arise as ${\rm g}_i(k)=k^{-d_i}
\bar{\rm g}_i(k)$ by rescaling (with the cutoff $k$) the original couplings 
$\bar{\rm g}_i$ with mass 
dimensions $d_i$, then $\mbox{\boldmath $\beta$}_i=-d_i{\rm g}_i+\cdots$ and 
$B_{ij}=-d_i\delta_{ij}+\cdots$ where the dots stand for the quantum
corrections. Ignoring them, $\theta_i=d_i+\cdots$, and $\Delta_{\rm UV}$
equals the number of positive $d_i$'s. Since adding derivatives or powers of
fields to a monomial in the action always lowers $d_i$, there can be at most
a finite number of positive $d_i$'s and, therefore, of negative eigenvalues
of ${\bf B}$. Thus, barring the presumably rather exotic possibility that the 
quantum corrections change
the signs of infinitely many elements in ${\bf B}$, the dimensionality of
${\mathcal S}_{\rm UV}$ is finite \cite{wein}.

We emphasize that in general the UV fixed point on which the above construction
is based, if it exists, has no reason to be of the simple Einstein-Hilbert 
form (\ref{l1}).
The initial point of the RG trajectory $\Gamma_{k\rightarrow\infty}$ is 
expected to contain many more invariants, both local (curvature polynomials)
and nonlocal ones. For this reason the asymptotic safety scenario is {\it not}
a quantization of General Relativity, and it cannot be compared in this respect
to the loop quantum gravity approach,
for instance. In a conventional field theory setting the functional 
$\Gamma_{k\rightarrow\infty}$ corresponds to the bare (or ``classical'') action
$S$ which usually can be chosen (almost) freely. It is one of the many 
attractive features of the asymptotic safety scenario that the bare action is
fixed by the theory itself and actually can be {\it computed}, namely
by searching for zeros of
$\vec{\mbox{\boldmath$\beta$}}$. In this respect it has, almost by 
construction, a degree of predictivity which cannot be reached by any scheme
trying to quantize a given classical action.

\section{RG flow of the effective average action}
\renewcommand{\theequation}{3.\arabic{equation}}
\setcounter{equation}{0}
\label{eaa}
During the past few years, the asymptotic safety scenario in
Quantum Einstein Gravity (QEG) has been mostly
investigated in the framework of the effective average action 
\cite{mr}-\cite{hier}, \cite{bagber}, a specific formulation of the 
Wilsonian RG which
originally was developed for theories in flat space \cite{avact,ym,avactrev}
and has been first applied to gravity in \cite{mr}.

Quite generally, the effective average action $\Gamma_k$ is a coarse grained 
free energy
functional that describes the behavior of the theory at the mass scale $k$. 
It contains the quantum effects of all fluctuations of the 
dynamical variables with momenta larger than $k$, but not of those 
with momenta smaller than $k$. As $k$ is decreased, an increasing number 
of degrees of freedom is integrated out. The method thus complies, at an 
intuitive level, with the coarse graining picture of the previous section. The 
successive averaging of the fluctuation variable is achieved by a 
$k$-dependent IR cutoff term 
$\Delta_kS$ which is added to the classical action in the standard Euclidean 
functional integral. This term gives a momentum dependent mass square 
${\mathcal R}_k(p^2)$ to the field modes with momentum $p$. It is designed
to vanish if $p^2\gg k^2$, but suppresses the contributions of the modes with
$p^2< k^2$ to the path integral.
When regarded as a function of $k$, $\Gamma_k$ describes a curve in theory
space that interpolates between the classical
action $S=\Gamma_{k\rightarrow\infty}$ and the conventional effective action
$\Gamma=\Gamma_{k=0}$. The change of $\Gamma_k$ induced by an infinitesimal
change of $k$ is described by a functional differential equation, the exact
RG equation. In a symbolic notation it reads
\begin{eqnarray}
\label{newRG}
k\,\partial_k\Gamma_k=\frac{1}{2}\;{\rm STr}\left[\left(\Gamma_k^{(2)}
+{\mathcal R}_k\right)^{-1}\,k\,\partial_k{\mathcal R}_k\right]\;.
\end{eqnarray}
For a detailed discussion of this equation we must refer to the literature
\cite{mr}. Suffice it to say that, expanding $\Gamma_k[g_{\mu\nu},\cdots]$ in 
terms of diffeomorphism invariant field
monomials $I_i[g_{\mu\nu},\cdots]$ with coefficients ${\rm g}_i(k)$, eq.
(\ref{newRG}) assumes the component form (\ref{rgrho}).  

In general it is impossible to find exact solutions to eq. (\ref{newRG})
and we are forced to rely upon approximations. A powerful
nonperturbative approximation scheme is the truncation of theory space
where the RG flow is projected onto a finite-dimensional subspace. In practice
one makes an ansatz for $\Gamma_k$ that
comprises only a few couplings and inserts it into the RG equation. This
leads to a, now finite, set of coupled differential equations of the form
(\ref{rgrho}). 

The simplest approximation one might try is the 
``Einstein-Hilbert truncation'' \cite{mr,oliver1} defined by the ansatz
\begin{eqnarray}
\label{in2}
\Gamma_k[g_{\mu\nu}]=\left(16\pi G_k\right)^{-1}\int d^dx\,\sqrt{g}\left\{
-R(g)+2\bar{\lambda}_k\right\}
\end{eqnarray}
It applies to a $d$-dimensional Euclidean spacetime and involves only the 
cosmological constant $\bar{\lambda}_k$ and the Newton
constant $G_k$ as running parameters. Inserting (\ref{in2}) into
the RG equation (\ref{newRG}) one obtains a set of two
$\mbox{\boldmath$\beta$}$-functions $(\mbox{\boldmath$\beta$}_\lambda,
\mbox{\boldmath$\beta$}_g)$ for the dimensionless cosmological constant 
$\lambda_k\equiv k^{-2}\bar{\lambda}_k$ and the dimensionless Newton constant
$g_k\equiv k^{d-2}G_k$, respectively. They describe a two-dimensional RG flow 
on the plane with coordinates ${\rm g}_1\equiv\lambda$ and ${\rm g}_2\equiv g$.
At a fixed point $(\lambda_*,g_*)$, both $\mbox{\boldmath$\beta$}$-functions
vanish simultaneously. In the Einstein-Hilbert truncation there
exists both a trivial Gaussian fixed point (GFP) at $\lambda_*=g_*=0$ and, 
quite remarkably, also a UV attractive NGFP at $(\lambda_*,g_*)\neq(0,0)$.
\begin{figure}[t]
\renewcommand{\baselinestretch}{1}
\epsfxsize=0.99 \textwidth
\centerline{\epsfbox{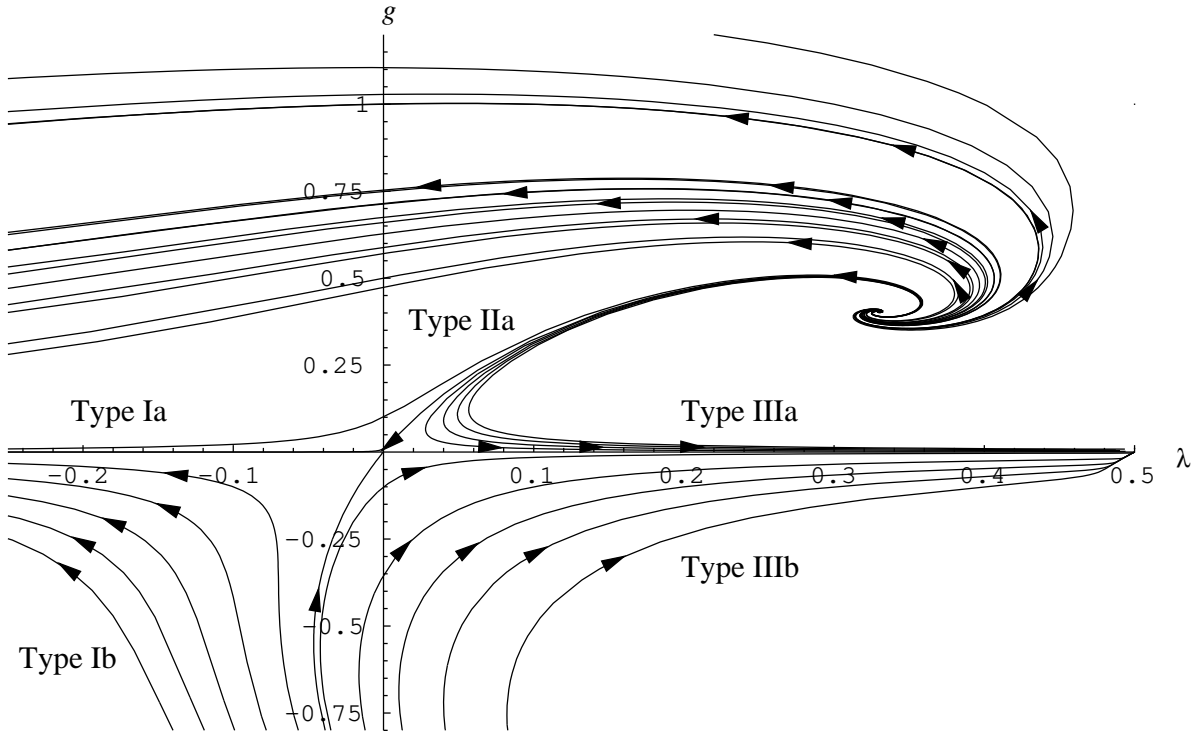}}
\parbox[c]{\textwidth}{\caption{\label{1.eins}{Part of theory space of the 
Einstein-Hilbert truncation with its RG flow. The arrows point in the direction
of decreasing values of $k$. The flow is dominated by a non-Gaussian 
fixed point in the first quadrant and a trivial one at the origin. (From 
\cite{frank1}.)}}}
\end{figure}

In Fig. \ref{1.eins} we show part of the $g$-$\lambda$ theory space and the
corresponding RG flow for $d=4$. The trajectories are obtained by numerically
integrating the differential equations $k\,\partial_k\lambda=\mbox{\boldmath
$\beta$}_\lambda(\lambda,g)$ and $k\,\partial_k g=\mbox{\boldmath
$\beta$}_g(\lambda,g)$. The arrows point in the direction of increasing coarse
graining, i.e. from the UV towards the IR. We observe that three types of
trajectories emanate from the NGFP: those of 
Type Ia (Type IIIa) run towards negative (positive) cosmological constants,
while the ``separatrix'', the unique trajectory (of Type IIa) crossing over
from the NGFP to the GFP, has a vanishing 
cosmological constant in the IR. The flow is defined on the half-plane
$\lambda<1/2$ only; it cannot be continued beyond $\lambda=1/2$ as the 
{\boldmath$\beta$}-functions become singular there. In fact, the 
Type IIIa-trajectories cannot be integrated down to $k=0$ within
the Einstein-Hilbert approximation. They terminate at a non-zero $k_{\rm term}$
where they run into the $\lambda=1/2-$singularity. Near $k_{\rm term}$ a more
general truncation is needed in order to continue the flow.

In Weinberg's original paper \cite{wein} the asymptotic safety idea was 
tested in 
$d=2+\epsilon$ dimensions where $0<\epsilon\ll 1$ was chosen so that the 
{\boldmath$\beta$}-functions (actually $\mbox{\boldmath$\beta$}_g$ only) could
be found by an $\epsilon$-expansion. Before the advent of the exact RG
equations no practical tool was known which would have allowed a 
nonperturbative calculation of the {\boldmath$\beta$}-functions in the
physically interesting case of $d=4$ spacetime dimensions. However, as we saw 
above, the
effective average action in the Einstein-Hilbert approximation does indeed 
predict the existence of a NGFP in a nonperturbative setting. It was first
analyzed in \cite{souma,oliver1,frank1}, and also first investigations of its
possible role in black hole physics \cite{bh} and cosmology 
\cite{cosmo1,cosmo2} were performed already. 

The detailed analyses of refs.
\cite{oliver1,frank1} demonstrated that the NGFP found has all the properties
necessary for asymptotic safety. In particular one has a pair of complex
conjugate critical exponents $\theta'\pm{\rm i}\,\theta''$ with $\theta'>0$,
implying that the NGFP, for $k\rightarrow\infty$, attracts all trajectories in
the half-plane $g>0$. (The lower half-plane $g<0$ is unphysical probably 
since it corresponds to a negative Newton constant.) Because of the 
nonvanishing imaginary part $\theta''\neq 0$, all trajectories spiral around
the NGFP before hitting it.

The question of crucial importance is whether the fixed point predicted by
the Einstein-Hilbert truncation actually approximates a fixed point in the
exact theory, or whether it is an artifact of the truncation. In refs.
\cite{oliver1,oliver2,frank1} evidence was found which, in our opinion,
strongly supports the hypothesis that there does indeed exist a non-Gaussian 
fixed point in the exact 4-dimensional theory, with exactly the properties 
required for the asymptotic safety scenario. In these investigations the
reliability of the Einstein-Hilbert truncation was tested
both by analyzing the cutoff scheme dependence within this truncation 
\cite{oliver1,frank1} and by generalizing the truncation ansatz itself
\cite{oliver2}. The idea behind the first method is as follows.

The cutoff operator ${\mathcal R}_k(p^2)$ is specified by a matrix in field 
space
and a ``shape function'' $R^{(0)}(p^2/k^2)$ which describes the details of how
the modes get suppressed in the IR when $p^2$ drops below $k^2$. We checked the
cutoff scheme dependence of the various quantities of interest both by looking
at their dependence on the function $R^{(0)}$ and comparing two different
matrix structures. Universal quantities are particularly important in this 
respect because, by definition, they are strictly cutoff scheme independent 
in the exact theory. Any truncation leads to a residual scheme dependence of 
these quantities, however. Its magnitude
is a natural indicator for the quality of the truncation \cite{kana}. Typical
examples of universal quantities are the critical exponents $\theta_I$. The
existence or nonexistence of a fixed point is also a universal, scheme 
independent feature, but its precise location in parameter space is scheme
dependent. Nevertheless it can be shown that, in $d=4$, the product $g_*
\lambda_*$ must be universal \cite{oliver1} while $g_*$ and $\lambda_*$ 
separately are not.

The detailed numerical analysis of the Einstein-Hilbert RG flow near the
NGFP \cite{oliver1,frank1} shows that the universal quantities, in particular
the product $g_*\,\lambda_*$, are indeed scheme independent at a quite
impressive level of accuracy. As the many numerical ``miracles'' which lead to
the almost perfect cancellation of the $R^{(0)}$-dependence would have no 
reason to occur if there was not a fixed point in the exact theory as an 
organizing principle, the results of this
analysis can be considered strong evidence in favor of a fixed point in the
exact, un-truncated theory.

The ultimate justification of any truncation is that when one adds 
further terms to it its physical predictions do not change significantly any 
more. As a first step towards testing the stability of the Einstein-Hilbert
truncation against the inclusion of other invariants \cite{oliver2} we took a 
$({\rm curvature})^2$-term into account:
\begin{eqnarray}
\label{p2.4}
\Gamma_k[g_{\mu\nu}]=\int d^dx\,\sqrt{g}\left\{\left(16\pi G_k\right)^{-1}
\left[-R(g)+2\bar{\lambda}_k\right]+\bar{\beta}_k\,R^2(g)\right\}
\end{eqnarray}
Inserting (\ref{p2.4}) into the functional RG equation yields a set of 
$\mbox{\boldmath$\beta$}$-functions 
$(\mbox{\boldmath$\beta$}_\lambda,\mbox{\boldmath$\beta$}_g,
\mbox{\boldmath$\beta$}_\beta)$ for the dimensionless couplings $\lambda_k$,
$g_k$ and $\beta_k\equiv k^{4-d}\bar{\beta}_k$. They describe the RG flow on
the three-dimensional $\lambda$-$g$-$\beta-$space. Despite the extreme 
algebraic complexity of the three {\boldmath$\beta$}-functions it was possible
to show \cite{oliver2,oliver3,oliver4} that they, too, admit a NGFP 
$(\lambda_*,g_*,\beta_*)$ with exactly the properties needed for asymptotic
safety. In particular it turned out to be UV attractive in all three 
directions. The value
of $\beta_*$ is extremely tiny, and close to the NGFP the projection of the
3-dimensional flow onto the $\lambda$-$g-$subspace is very well described by
the Einstein-Hilbert truncation which ignores the third direction from the
outset. The $\lambda_*$- and $g_*$-values and the critical exponents related to
the flow in the $\lambda$-$g-$subspace, as predicted by the 3-dimensional
truncation, agree almost perfectly  with those from the Einstein-Hilbert
approximation. Analyzing the scheme dependence of the universal quantities
one finds again a highly remarkable $R^{(0)}$-independence $-$ which is truly
amazing if one visualizes the huge amount of nontrivial numerical 
compensations and cancellations among several dozens of $R^{(0)}$-dependent
terms which is necessary to make $g_*\,\lambda_*$, say, approximately 
independent of the shape function $R^{(0)}$.

On the basis of these results we believe that the non-Gaussian fixed
point occuring in the Einstein-Hilbert truncation is very unlikely to be an
artifact of this truncation but rather may be considered the projection of a 
NGFP in the exact theory. The fixed point
and all its qualitative properties are stable against variations of the cutoff
and the inclusion of a further invariant in the truncation. It is particularly
remarkable that within the scheme dependence the additional $R^2$-term has
essentially no impact on the fixed point. These are certainly very nontrivial
indications supporting the conjecture that 4-dimensional QEG
indeed possesses a RG fixed point with the properties needed for its 
nonperturbative renormalizability. 

This view is further supported by two conceptually independent investigations.
In ref. \cite{prop} a proper time renormalization group equation rather than
the flow equation of the average action has been used, and again a suitable 
NGFP was found.
This framework is conceptually somewhat simpler than that of the effective
average action; it amounts to an RG-improved 1-loop calculation with an IR
cutoff. Furthermore, in refs. \cite{max} the functional integral over the 
subsector of metrics admitting two Killing vectors has been performed 
{\it exactly}, and again
a NGFP was found, this time in a setting and an approximation which is 
{\it very} different from that of
the truncated $\Gamma_k$-flows. As for the inclusion of matter fields, both
in the average action \cite{perper1,perper2,perini,essential} and the symmetry
reduction approach \cite{max} a suitable NGFP has been established for a broad
class of matter systems.

\section{Scale dependent metrics and the\\ resolution function \boldmath
$\ell(k)$}
\renewcommand{\theequation}{4.\arabic{equation}}
\setcounter{equation}{0}
\label{sdm}
In the following we take the existence of a suitable NGFP on the full theory
space for granted and explore some of the properties of asymptotic safety,
in particular we try to gain some understanding of what a ``quantum spacetime''
is like. Unless stated otherwise we consider pure Euclidean gravity in $d=4$.

The running effective average action $\Gamma_k[g_{\mu\nu}]$ defines an infinite
set of effective field theories, valid near the scale $k$ 
which we may vary between $k=0$ and $k=\infty$. Intuitively speaking, 
the solution $\big<g_{\mu\nu}\big>_k$ of the scale dependent field equation 
\begin{eqnarray}
\label{fe}
\frac{\delta\Gamma_k}{\delta g_{\mu\nu}(x)}[\big<g\big>_k]=0
\end{eqnarray}
can be interpreted as the metric averaged over (Euclidean) spacetime volumes
of a linear extension $\ell$ which typically is of the order of $1/k$. Knowing
the scale dependence of $\Gamma_k$, i.e. the renormalization group trajectory
$k\mapsto\Gamma_k$, we can derive the running effective Einstein equations
(\ref{fe}) for any $k$ and, after fixing appropriate boundary conditions and
symmetry requirements, follow their solution 
$\big<g_{\mu\nu}\big>_k$ from $k=\infty$ to $k=0$.

The infinitely many equations
of (\ref{fe}), one for each scale $k$, are valid {\it simultaneously}. They
all refer {\it to the same} physical system, the ``quantum spacetime''. They
describe its effective metric structure on different length scales. An 
observer using
a ``microscope'' with a resolution $\approx k^{-1}$ will perceive the universe
to be a Riemannian manifold with metric $\big<g_{\mu\nu}\big>_k$. At every 
fixed $k$, $\big<g_{\mu\nu}\big>_k$ is a smooth classical metric. But since
the quantum spacetime is characterized by the infinity of metrics 
$\{\big<g_{\mu\nu}\big>_k|k=0,\cdots,\infty\}$ it can acquire very nonclassical
and in particular fractal features. In fact, every proper distance calculated 
from $\big<g_{\mu\nu}\big>_k$ is unavoidably scale dependent. This phenomenon 
is familiar from fractal geometry, a famous example being the coast line of
England whose length depends on the size of the yardstick used to measure
it \cite{mandel}.

Let us describe more precisely what it means to ``average'' over Euclidean
spacetime volumes. The quantity we can freely tune is the IR cutoff scale $k$,
and the ``resolving power'' of the microscope, henceforth denoted $\ell$, is an
a priori unknown function of $k$. (In flat space, $\ell\approx \pi/k$.)
In order to understand the relationship between $\ell$ and $k$ we must recall
some more steps from the construction of $\Gamma_k[g_{\mu\nu}]$ in ref. 
\cite{mr}.

The effective average action is obtained by introducing an IR cutoff into the
path-integral over all metrics, gauge fixed by means of a background gauge
fixing condition. Even without a cutoff the resulting effective action 
$\Gamma[g_{\mu\nu};\bar{g}_{\mu\nu}]$ depends on two metrics, the expectation
value of the quantum field, $g_{\mu\nu}$, and the background field 
$\bar{g}_{\mu\nu}$. This is a standard technique, and it is well known 
\cite{back} that the functional 
$\Gamma[g_{\mu\nu}]\equiv\Gamma[g_{\mu\nu};\bar{g}_{\mu\nu}=g_{\mu\nu}]$ 
obtained by equating the two
metrics can be used to generate the 1PI Green's functions of the theory.

(We emphasize, however, that the average action method is manifestly {\it
background independent} despite the temporary use of $\bar{g}_{\mu\nu}$ at an
intermediate level. At no stage in the derivation of the 
{\boldmath$\beta$}-functions it is necessary to assign a concrete metric to
$\bar{g}_{\mu\nu}$, such as $\bar{g}_{\mu\nu}=\eta_{\mu\nu}$ in standard
perturbation theory, say. The RG flow, i.e. the vector field 
$\vec{\mbox{\boldmath$\beta$}}$, on the theory space of diffeomorphism 
invariant action functionals depending on $g_{\mu\nu}$ and $\bar{g}_{\mu\nu}$
is a highly universal object: it neither depends on any specific metric, nor on
any specific action.)

The IR cutoff of the average action is implemented by first expressing the
functional integral over all metric fluctuations in terms of eigenmodes of 
$\bar{D}^2$, the covariant Laplacian formed with the aid of the background 
metric $\bar{g}_{\mu\nu}$. Then the suppression term $\Delta_k S$ is 
introduced which damps the
contribution of all $-\bar{D}^2$-modes with eigenvalues smaller than $k^2$.
Coupling the dynamical fields to sources and Legendre-transforming leads to 
the scale dependent
functional $\Gamma_k[g_{\mu\nu};\bar{g}_{\mu\nu}]$, and the action with one
argument again obtains by equating the two metrics:
$\Gamma_k[g_{\mu\nu}]\equiv\Gamma_k[g_{\mu\nu};\bar{g}_{\mu\nu}=g_{\mu\nu}]$.
It is this action which appears in (\ref{fe}). Because of the identification
of the two metrics we see that, in a sense, it is the eigenmodes of
$\bar{D}^2=D^2$,
constructed from the argument of $\Gamma_k[g]$, which are cut off at $k^2$.

This last observation is essential for the following algorithm \cite{ym,jan}
for the reconstruction of the averaging scale $\ell$ from the cutoff $k$. The 
input data is the set of metrics characterizing a quantum manifold, 
$\{\big<g_{\mu\nu}\big>_k\}$. The idea is to deduce the relation 
$\ell = \ell(k)$ from the spectral properties of the scale dependent 
Laplacian $\Delta(k) \equiv D^2(\big<g_{\mu\nu}\big>_k)$ built with the 
solution of the effective field equation. More precisely, for every fixed 
value of $k$, one solves the eigenvalue problem of $-\Delta(k)$ and studies  
the properties of the special eigenfunctions whose eigenvalue is $k^2$, or 
nearest to $k^2$ in the case of a discrete spectrum. We shall refer to an
eigenmode of $-\Delta(k)$ whose eigenvalue is (approximately) the square of the
cutoff $k$ as a ``cutoff mode" (COM) and denote the set of all COMs by 
{\sf COM}($k$).

If we ignore the $k$-dependence of $\Delta(k)$ for a moment (as it would be
appropriate for matter theories in flat space) the COMs are, for a sharp 
cutoff,
precisely the last modes integrated out when lowering the cutoff, since the 
suppression term in the path integral cuts out all modes of the metric
fluctuation with eigenvalue smaller than $k^2$.

For a non-gauge theory in flat space the coarse graining or averaging of 
fields is a well defined procedure, based upon ordinary Fourier analysis,
and one finds that in this case the length $\ell$
is essentially the wave length of the last modes integrated out, the COMs.

This observation motivates the following definition of $\ell$ in quantum 
gravity. We determine the COMs of $-\Delta(k)$, analyze how fast these 
eigenfunctions vary on spacetime, and read off a typical coordinate distance 
$\Delta x^\mu$ characterizing the scale on which they vary. For an oscillatory
COM, for example, $\Delta x^\mu$ would correspond to an oscillation period. 
(In general there is a certain freedom in the precise identification of the 
$\Delta x^\mu$ belonging to a specific cutoff mode. This ambiguity can be 
resolved by refining the definition of $\Delta x^\mu$ on a case-by-case basis 
only.) Finally we use the metric $\big<g_{\mu\nu}\big>_k$ itself in order to 
convert $\Delta x^\mu$ to a proper length. This proper length, by definition, 
is $\ell$. Repeating the above steps for all values of $k$, we end up with a 
function $\ell =\ell(k)$. In general one will find that $\ell$ depends on the 
position on the manifold as well as on the direction of $\Delta x^\mu$. 

Applying the above algorithm on a non-dynamical flat spacetime one recovers the
expected result $\ell(k)=\pi/k$. In ref. \cite{jan} a specific example of a
QEG spacetime has been constructed, the quantum $S^4$, which is an ordinary
4-sphere at every fixed scale, with a $k$-dependent radius, though. In this 
case, too, the resolution function was found to be $\ell(k)=\pi/k$.

Thus the construction and interpretation of a QEG spacetime proceeds, in a
nutshell, as follows. We start from a fixed RG trajectory
$k\mapsto\Gamma_k$, derive its effective field equations at each $k$, and solve
them. The resulting quantum mechanical counterpart of a classical spacetime is 
equipped with the infinity of Riemannian metrics $\{\big<g_{\mu\nu}\big>_k\big|
k=0,\cdots,\infty\}$ where the parameter $k$ is only a book keeping device 
a priori. It can be given a physical interpretation by relating it
to the COM length scale characterizing the averaging procedure: One constructs
the Laplacian $-D^2(\big<g_{\mu\nu}\big>_k)$, diagonalizes it, looks how 
rapidly its $k^2$-eigenfunction varies, and ``measures'' the length of typical 
variations with the metric $\big<g_{\mu\nu}\big>_k$ itself. In the ideal case
one can solve the resulting $\ell=\ell(k)$ for $k=k(\ell)$ and reinterprete the
metric $\big<g_{\mu\nu}\big>_k$ as referring to a microscope with a known
position and direction dependent resolving power $\ell$. The price we have to 
pay for the background independence is that we cannot freely choose $\ell$ 
directly but rather $k$ only.

\section{Microscopic structure of the QEG spacetimes}
\renewcommand{\theequation}{5.\arabic{equation}}
\setcounter{equation}{0}
\label{msqeg}
One of the intriguing conclusions we reached in refs. 
\cite{oliver1,oliver2} was that the QEG spacetimes are fractals and that their
effective dimensionality is
scale dependent. It equals 4 at macroscopic distances ($\ell\gg\ell_{\rm Pl}$)
but, near $\ell\approx\ell_{\rm Pl}$, it gets dynamically reduced to the value
2. For $\ell\ll\ell_{\rm Pl}$ spacetime is, in a precise sense \cite{oliver1},
a 2-dimensional fractal.

In ref. \cite{cosmo1} the specific form of the graviton propagator on this 
fractal was applied in a cosmological context. It was argued that it gives rise
to a Harrison-Zeldovich spectrum of primordial geometry fluctuations, perhaps
responsible for the CMBR spectrum observed today.
(In refs. \cite{bh,cosmo1,cosmo2}, \cite{elo}-\cite{mof} various types of 
``RG improvements'' were used to explore possible physical manifestations of 
the scale dependence of the gravitational parameters.)

A priori there exist several plausible definitions of a fractal dimensionality
of spacetime. In our original argument \cite{oliver1} we used the one based
upon the anomalous dimension $\eta_N$ at the NGFP. We shall review this 
argument in the rest of this section. Then, in Section \ref{specdimen}, we
evaluate the spectral dimension for the QEG spacetimes \cite{oliverfrac} and
demonstrate that it displays the same dimensional reduction $4\rightarrow
2$ as the one based upon $\eta_N$. The spectral dimension has also been
determined in Monte Carlo simulations of causal (i.e. Lorentzian) dynamical
triangulations \cite{amb}-\cite{ajl34} and it will be interesting to compare
the results.

For simplicity we use the Einstein-Hilbert truncation to start with, and we
consider spacetimes with classical dimensionality $d=4$. The corresponding RG
trajectories are shown in Fig.~\ref{1.eins}. For $k\rightarrow\infty$, all of
them approach the NGFP $(\lambda_*,g_*)$ so that the dimensionful quantities
run according to
\begin{eqnarray}
\label{asymrun}
G_k\approx g_*/k^2\;\;\;\;,\hspace{2cm}\bar{\lambda}_k\approx\lambda_*\,k^2
\end{eqnarray}
The behavior (\ref{asymrun}) is realized in the asymptotic scaling regime
$k\gg m_{\rm Pl}$. Near $k= m_{\rm Pl}$ the trajectories cross over towards the
GFP. Since we are interested only in the limiting cases of very small
and very large distances the following caricature of a RG trajectory will be
sufficient. We assume that $G_k$ and $\bar{\lambda}_k$ behave as in 
(\ref{asymrun}) for $k\gg m_{\rm Pl}$, and that they assume constant values
for $k\ll m_{\rm Pl}$. The precise interpolation between the two regimes could
be obtained numerically \cite{frank1} but will not be needed here.

The argument of ref. \cite{oliver2} concerning the fractal nature of the QEG
spacetimes is as follows. Within the Einstein-Hilbert truncation of theory
space, the effective field equations (\ref{fe}) happen to coincide with the 
ordinary Einstein equation, but with $G_k$ and $\bar{\lambda}_k$ replacing the
classical constants. Without matter,
\begin{eqnarray}
\label{einsteq}
R_{\mu\nu}(\big<g\big>_k)
&=&\bar{\lambda}_k\,\big<g_{\mu\nu}\big>_k
\end{eqnarray}
Since in absence of dimensionful constants of integration
$\bar{\lambda}_k$ is the only quantity in this equation which sets a 
scale, every solution to (\ref{einsteq}) has a typical radius of curvature 
$r_c(k)\propto 1/\sqrt{\bar{\lambda}_k}$. (For instance, the
$S^4$-solution has the radius $r_c=\sqrt{3/\bar{\lambda}_k}$.) 
If we want to explore the spacetime structure at a fixed length scale $\ell$ 
we should use the action $\Gamma_k[g_{\mu\nu}]$ at $k\approx\pi/\ell$ because 
with 
this functional a tree level analysis is sufficient to describe the essential 
physics at this scale, including the relevant quantum effects. Hence, when we 
observe the spacetime with a microscope of resolution $\ell$, we will see an 
average radius of curvature given by 
$r_c(\ell)\equiv r_c(k=\pi/\ell)$. Once $\ell$ is
smaller than the Planck length $\ell_{\rm Pl}\equiv m_{\rm Pl}^{-1}$
we are in the fixed point regime where $\bar{\lambda}_k\propto k^2$ so that 
$r_c(k)\propto 1/k$, or
\begin{eqnarray}
\label{radius}
r_c(\ell)\propto\ell
\end{eqnarray}
Thus, when we look at the structure of spacetime with a microscope of 
resolution $\ell\ll\ell_{\rm Pl}$, the average radius 
of curvature which we measure is proportional to the resolution 
itself. If we want to probe finer details and decrease $\ell$ we automatically
decrease $r_c$ and hence {\it in}crease the average curvature. Spacetime seems
to be more strongly curved at small distances than at larger ones. The 
scale-free relation (\ref{radius}) suggests that at distances below the Planck
length the QEG spacetime is a special kind of fractal with a self-similar 
structure. It has no intrinsic scale because in the fractal regime, i.e. when 
the RG trajectory is still close to the NGFP, the parameters which usually
set the scales of the gravitational interaction, $G$ and $\bar{\lambda}$, are 
not yet ``frozen out''. This happens only later on, somewhere half way between
the NGFP and the GFP, at a scale of the order of 
$m_{\rm Pl}$.
Below this scale, $G_k $ and $\bar{\lambda}_k$ stop running and, as a result,
$r_c(k)$ becomes independent of $k$ so that $r_c(\ell)={\rm const}$ for 
$\ell\gg\ell_{\rm Pl}$. In this regime $\big<g_{\mu\nu}\big>_k$ is 
$k$-independent, indicating that the macroscopic spacetime is describable by a
single smooth Riemannian manifold.

The above argument made essential use of the proportionality $\ell\propto 1/k$.
In the fixed point regime it follows trivially from the fact that there exist 
no other relevant dimensionful parameters so that $1/k$ is the only length 
scale one can form. The algorithm for the determination of $\ell(k)$ described
above yields the same answer.

It is easy to make the $k$-dependence of $\big<g_{\mu\nu}\big>_k$ explicit.
Picking an arbitrary reference scale $k_0$ we rewrite (\ref{einsteq}) as
$[\bar{\lambda}_{k_0}/\bar{\lambda}_k]\,R^\mu_{\;\;\nu}(\big<g\big>_k)
=\bar{\lambda}_{k_0}\,\delta^\mu_\nu$. Since $R^\mu_{\;\;\nu}(c\,g)=c^{-1}\,
R^\mu_{\;\;\nu}(g)$ for any constant $c>0$, the average metric and its inverse
scale as
\begin{eqnarray}
\label{metscale}
\big<g_{\mu\nu}(x)\big>_k&=&[\bar{\lambda}_{k_0}/\bar{\lambda}_k]\,
\big<g_{\mu\nu}(x)\big>_{k_0}
\end{eqnarray}
\begin{eqnarray}
\label{invmetscale}
\big<g^{\mu\nu}(x)\big>_k&=&[\bar{\lambda}_k/\bar{\lambda}_{k_0}]\,
\big<g^{\mu\nu}(x)\big>_{k_0}
\end{eqnarray}
These relations are valid provided the family of solutions considered exists
for all scales between $k_0$ and $k$, and $\bar{\lambda}_k$ has the
same sign always.

As we discussed in ref. \cite{oliver1} the QEG spacetime has an effective
dimensionality which is $k$-dependent and hence noninteger in general. The
discussion was based upon the anomalous dimension $\eta_N$
of the operator $\int\sqrt{g}\,R$. It is defined as $\eta_N\equiv-k\,\partial_k
\ln Z_{Nk}$ where $Z_{Nk}\propto 1/G_k$ is the wavefunction renormalization
of the metric \cite{mr}. In a sense which we shall make more precise in a 
moment, the effective dimensionality of spacetime equals $4+\eta_N$. The RG
trajectories of the Einstein-Hilbert truncation (within its domain of validity)
have $\eta_N\approx 0$ for $k\rightarrow 0$\footnote{In the case of type IIIa
trajectories \cite{frank1,h3} the macroscopic $k$-value is still far above
$k_{\rm term}$, i.e. in the ``GR regime'' described in \cite{h3}.}
and $\eta_N\approx -2$ for 
$k\rightarrow\infty$, the smooth change by two units occuring near 
$k\approx m_{\rm Pl}$. As a consequence, the effective dimensionality is 4 for
$\ell\gg\ell_{\rm Pl}$ and 2 for $\ell\ll\ell_{\rm Pl}$.

The UV fixed point has an anomalous dimension $\eta\equiv\eta_N(
\lambda_*,g_*)=-2$. We can use this information in order to determine the 
momentum dependence of the dressed graviton propagator for momenta $p^2\gg
m_{\rm Pl}^2$. Expanding the $\Gamma_k$ of (\ref{in2}) about flat space
and omitting the standard tensor structures we find the inverse propagator
$\widetilde{\mathcal G}_k(p)^{-1}\propto Z_N(k)\,p^2$. The conventional dressed
propagator $\widetilde{\mathcal G}(p)$, the one contained in 
$\Gamma\equiv\Gamma_{k=0}$, obtains from the exact $\widetilde{\mathcal G}_k$ 
by taking the limit $k\rightarrow 0$.
For $p^2>k^2\gg m_{\rm Pl}^2$ the actual cutoff scale is the physical momentum
$p^2$ itself\footnote{See Section 1 of ref. \cite{h1} for a detailed discussion
of ``decoupling'' phenomena of this kind.}
so that the $k$-evolution of $\widetilde{\mathcal G}_k(p)$ stops at the 
threshold $k=\sqrt{p^2}$. Therefore
\begin{eqnarray}
\label{gp1}
\widetilde{\mathcal G}(p)^{-1}\propto\;Z_N\left(k=\sqrt{p^2}\right)\,p^2
\propto\;(p^2)^{1-\frac{\eta}{2}}
\end{eqnarray}
because $Z_N(k)\propto k^{-\eta}$ when $\eta\equiv-\partial_t\ln Z_N$ is 
(approximately) constant. In $d$ dimensions, and for $\eta\neq 2-d$, the
Fourier transform of $\widetilde{\mathcal G}(p)\propto 1/(p^2)^{1-\eta/2}$ 
yields the following propagator in position space:
\begin{eqnarray}
\label{gp2}
{\mathcal G}(x;y)\propto\;\frac{1}{\left|x-y\right|^{d-2+\eta}}\;.
\end{eqnarray}
This form of the propagator is well known from the theory of critical 
phenomena, for instance. (In the latter case it applies to large distances.)
Eq. (\ref{gp2}) is not valid directly at the NGFP. For $d=4$ and $\eta=-2$
the dressed propagator is $\widetilde{\mathcal G}(p)=1/p^4$ which has the 
following representation in position space:
\begin{eqnarray}
\label{gp3}
{\mathcal G}(x;y)=-\frac{1}{8\pi^2}\,\ln\left(\mu\left|x-y\right|\right)\;.
\end{eqnarray}
Here $\mu$ is an arbitrary constant with the dimension of a mass. Obviously
(\ref{gp3}) has the same form as a $1/p^2$-propagator in 2 dimensions.

Slightly away from the NGFP, before other physical scales intervene, the 
propagator is of the familiar type (\ref{gp2}) which shows that the quantity 
$\eta_N$
has the standard interpretation of an anomalous dimension in the sense that
fluctuation effects modify the decay properties of ${\mathcal G}$ so as to 
correspond to a spacetime of effective dimensionality $4+\eta_N$.

Thus the properties of the RG trajectories imply the following ``dimensional
reduction'': Spacetime, probed by a ``graviton'' with $p^2\ll m_{\rm Pl}^2$ is
4-dimensional, but it appears to be 2-dimensional for a graviton with 
$p^2\gg m_{\rm Pl}^2$ \cite{oliver1}.

It is interesting to note that in $d$ classical dimensions, where the 
macroscopic spacetime is $d$-dimensional, the anomalous dimension at the
fixed point is $\eta=2-d$. Therefore, for any $d$, the dimensionality of the
fractal as implied by $\eta_N$ is $d+\eta=2$ \cite{oliver1,oliver2}.

\section{The spectral dimension}
\renewcommand{\theequation}{6.\arabic{equation}}
\setcounter{equation}{0}
\label{specdimen}
Next we turn to the spectral dimension ${\mathcal D}_{\rm s}$ of 
the QEG spacetimes. This particular definition of a fractal dimension is 
borrowed from the theory of diffusion processes on fractals \cite{avra} and 
is easily adapted to the
quantum gravity context \cite{nino,ajl34}. In particular it has been used in
the Monte Carlo studies mentioned above.

Let us study the diffusion of a scalar test particle on a $d$-dimensional 
classical Euclidean manifold with a fixed smooth metric $g_{\mu\nu}(x)$.
The corresponding heat-kernel $K_g(x,x';T)$ giving the probability for the
particle to diffuse from $x'$ to $x$ during the fictitious diffusion time $T$
satisfies the heat equation 
$\partial_T K_g(x,x';T)=\Delta_g K_g(x,x';T)$
where $\Delta_g\equiv D^2$ denotes the scalar Laplacian: 
$\Delta_g\phi\equiv g^{-1/2}\,\partial_\mu(g^{1/2}\,g^{\mu\nu}\,\partial_\nu
\phi)$. The heat-kernel is a matrix element of the operator 
$\exp(T\,\Delta_g)$. In the random walk picture its trace per unit volume,
$P_g(T)\equiv\, V^{-1}\,\int d^dx\,\sqrt{g(x)}\,K_g(x,x;T)\,\equiv\, 
V^{-1}\,{\rm Tr}\,\exp(T\,\Delta_g)$,
has the interpretation of an average return probability. (Here $V\equiv\int
d^dx\,\sqrt{g}$ denotes the total volume.) It is well known that $P_g$
possesses an asymptotic expansion (for $T\rightarrow 0$) of the form
$P_g(T)=(4\pi T)^{-d/2}\sum_{n=0}^\infty A_n\,T^n$. For an infinite flat
space, for instance, it reads $P_g(T)=(4\pi T)^{-d/2}$ for all $T$. Thus,
knowing the function $P_g$, one can recover the dimensionality of the target
manifold as the $T$-independent logarithmic derivative
$d=-2\,d\ln P_g(T)/d\ln T$.
This formula can also be used for curved spaces and spaces with finite volume
$V$ provided $T$ is not taken too large \cite{ajl34}.

In QEG where we functionally integrate over all metrics it is
natural to replace $P_g(T)$ by its expectation value. Symbolically,
$P(T)\equiv\big<P_\gamma(T)\big>$
where $\gamma_{\mu\nu}$ denotes the microscopic metric (integration variable)
and the expectation value is with respect to the ordinary path integral
(without IR cutoff) containing the fixed point action.
Given $P(T)$, we define the spectral dimension of the quantum
spacetime in analogy with the classical formula:
\begin{eqnarray}
\label{specdim}
{\mathcal D}_{\rm s}=-2\frac{d\ln P(T)}{d\ln T}
\end{eqnarray}

Let us now evaluate (\ref{specdim}) using the average
action method. The fictitious diffusion process takes place on a ``manifold''
which, at every fixed scale, is described by a smooth Riemannian metric
$\big<g_{\mu\nu}\big>_k$. While the situation appears to be classical at fixed
$k$, nonclassical features emerge in the regime with nontrivial RG running
since there the metric depends on the scale at which the spacetime structure
is probed. 

The nonclassical features are encoded in the properties of the diffusion
operator. Denoting the covariant Laplacians corresponding to the 
metrics $\big<g_{\mu\nu}\big>_k$ and $\big<g_{\mu\nu}\big>_{k_0}$ by
$\Delta(k)$ and $\Delta(k_0)$, respectively, eqs. (\ref{metscale}) and 
(\ref{invmetscale}) imply that they are related by
\begin{eqnarray}
\label{opscale0}
\Delta(k)&=&[\bar{\lambda}_k/\bar{\lambda}_{k_0}]\,\Delta(k_0)
\end{eqnarray}
When $k,k_0\gg m_{\rm Pl}$ we have, for example, 
\begin{eqnarray}
\label{opscale}
\Delta(k)&=&(k/k_0)^2\,\Delta(k_0)
\end{eqnarray}

Recalling that the average action $\Gamma_k$ defines an effective field 
theory at the scale $k$ we have that $\big<{\mathcal O}(\gamma_{\mu\nu})\big>
\approx{\mathcal O}(\big<g_{\mu\nu}\big>_k)$ if the operator ${\mathcal O}$ 
involves typical covariant momenta of the order $k$ and 
$\big<g_{\mu\nu}\big>_k$ solves eq. (\ref{fe}). In the following we exploit 
this relationship for the RHS of the diffusion equation, 
${\mathcal O}=\Delta_\gamma\,K_\gamma(x,x';T)$.
It is crucial here to correctly identify the relevant scale $k$.

If the diffusion process  involves only a small interval of 
scales near $k$ over which $\bar{\lambda}_k$ does not change much the 
corresponding heat equation must contain the $\Delta(k)$ for this specific, 
fixed value of $k$:
\begin{eqnarray}
\label{heateqk}
\partial_T K(x,x';T)&=&\Delta(k) K(x,x';T)
\end{eqnarray}
Denoting the eigenvalues of $-\Delta(k_0)$ by ${\mathcal E}_n$ and the 
corresponding eigenfunctions by $\phi_n$, this equation is solved by
\begin{eqnarray}
\label{kernexp}
K(x,x';T)&=&\sum\limits_n\phi_n(x)\,\phi_n(x')\,\exp\bigg(-F(k^2)\,
{\mathcal E}_n\,T\bigg)
\end{eqnarray}
Here we introduced the convenient notation $F(k^2)\equiv\bar{\lambda}_k/
\bar{\lambda}_{k_0}$. Knowing this propagation kernel we can time-evolve any
initial probability distribution $p(x;0)$ according to
$p(x;T)=\int d^4x'\,\sqrt{g_0(x')}\,K(x,x';T)\,p(x';0)$ with $g_0$ the 
determinant of $\big<g_{\mu\nu}\big>_{k_0}$. If the initial distribution has 
an eigenfunction expansion of the form $p(x;0)=\sum_n C_n\,\phi_n(x)$ we
obtain
\begin{eqnarray}
\label{probexp}
p(x;T)&=&\sum_n C_n\,\phi_n(x)\,\exp\bigg(-F(k^2)\,{\mathcal E}_n\,T\bigg)
\end{eqnarray}
If the $C_n$'s are significantly different from zero only for a single
eigenvalue ${\mathcal E}_N$, we are dealing with a single-scale problem. In the
usual spirit of effective field theories we would then identify 
$k^2={\mathcal E}_N$ as the relevant scale at which the running couplings are 
to be evaluated.
However, in general the $C_n$'s are different from zero over a wide range of
eigenvalues. In this case we face a multiscale problem where different modes
$\phi_n$ probe the spacetime on different length scales.

If $\Delta(k_0)$ corresponds to flat space, say, the eigenfunctions $\phi_n
\equiv\phi_p$ are plane waves with momentum $p^\mu$, and they resolve
structures on a length scale $\ell$ of order $\pi/|p|$. Hence, in terms of the
eigenvalue ${\mathcal E}_n\equiv{\mathcal E}_p=p^2$ the resolution is 
$\ell\approx
\pi/\sqrt{{\mathcal E}_n}$. This suggests that when the manifold is probed by a
mode with eigenvalue ${\mathcal E}_n$ it ``sees'' the metric 
$\big<g_{\mu\nu}\big>_k$ for the scale $k=\sqrt{{\mathcal E}_n}$. Actually the
identification $k=\sqrt{{\mathcal E}_n}$ is correct also for curved space 
since, in the construction of $\Gamma_k$, the parameter $k$ is introduced 
precisely as a cutoff in the spectrum of the covariant Laplacian. 

Therefore we conclude that under the spectral sum of (\ref{probexp}) we must
use the scale $k^2={\mathcal E}_n$ which depends explicitly on the resolving 
power of the
corresponding mode. Likewise, in eq. (\ref{kernexp}), $F(k^2)$ is to be
interpreted as $F({\mathcal E}_n)$. Thus we obtain the traced propagation 
kernel
\begin{eqnarray}
\label{trpropk}
P(T)&=&V^{-1}\;\sum_n \exp\bigg[-F({\mathcal E}_n)\,{\mathcal E}_n\,T\bigg]
\nonumber\\
&=&V^{-1}\;{\rm Tr}\,\exp\bigg[F\Big(-\Delta(k_0)\Big)\,\Delta(k_0)\,T\bigg]
\end{eqnarray}

It is convenient to choose $k_0$ as a macroscopic scale in a regime where there
are no strong RG effects any more.

Furthermore, let us assume for a moment that at $k_0$ the cosmological
constant is tiny, $\bar{\lambda}_{k_0}\approx 0$, so that $\big<g_{\mu\nu}
\big>_{k_0}$ is an approximately flat metric. In this case the trace in eq.
(\ref{trpropk}) is easily evaluated in a plane wave basis:
\begin{eqnarray}
\label{trplane}
P(T)=\int\frac{d^4p}{(2\pi)^4}\,\exp\left[-p^2\,F(p^2)\,T\right]
\end{eqnarray}
The $T$-dependence of (\ref{trplane}) determines the fractal dimensionality of
spacetime via (\ref{specdim}). In the limits $T\rightarrow\infty$ and 
$T\rightarrow 0$ where the random walks probe very large and small distances,
respectively, we obtain the dimensionalities corresponding to the largest
and smallest length scales possible. The limits $T\rightarrow\infty$ and 
$T\rightarrow 0$ of $P(T)$ are determined by the behavior of $F(p^2)\equiv
\bar{\lambda}(k=\sqrt{p^2})/\bar{\lambda}_{k_0}$ for $p^2\rightarrow 0$ and
$p^2\rightarrow\infty$, respectively.

For a RG trajectory where the renormalization effects stop below some threshold
we have $F(p^2\rightarrow 0)=1$. In this case (\ref{trplane}) yields
$P(T)\propto 1/T^2$, and we conclude that the macroscopic spectral dimension
is ${\mathcal D}_{\rm s}=4$.

In the fixed point regime we have $\bar{\lambda}_k\propto k^2$, and therefore
$F(p^2)\propto p^2$. As a result, the exponent in (\ref{trplane}) is 
proportional to $p^4$ now. This implies the $T\rightarrow 0-$behavior
$P(T)\propto 1/T$. It corresponds to the spectral dimension 
${\mathcal D}_{\rm s}=2$.

This result holds for all RG trajectories since only the fixed point 
properties were used. In particular it is independent of $\bar{\lambda}_{k_0}$
on macroscopic scales. Indeed, the above assumption that $\big<g_{\mu\nu}
\big>_{k_0}$ is flat was not necessary for obtaining ${\mathcal D}_{\rm s}=2$.
This follows from the fact that even for a curved metric the spectral sum
(\ref{trpropk}) can be represented by an Euler-Mac Laurin series which always
implies (\ref{trplane}) as the leading term for $T\rightarrow 0$.

Thus we may conclude that on very large and very small length scales the
spectral dimensions of the QEG spacetimes are
\begin{eqnarray}
\label{qegspecdims}
{\mathcal D}_{\rm s}(T\rightarrow\infty)&=&4\nonumber\\
{\mathcal D}_{\rm s}(T\rightarrow 0)&=&2
\end{eqnarray}

The dimensionality of the fractal at sub-Planckian distances is
found to be 2 again, as in the first argument based upon $\eta_N$.
Remarkably, the equality of $4+\eta$ and ${\mathcal D}_{\rm s}$
is a special feature of 4 classical dimensions. Generalizing for $d$
classical dimensions, the fixed point running of Newton's constant becomes
$G_k\propto k^{2-d}$ with a dimension-dependent exponent, while 
$\bar{\lambda}_k\propto k^2$ continues to have a quadratic $k$-dependence. As
a result, the $\widetilde{{\mathcal G}}(k)$ of eq. (\ref{gp1}) is proportional
to $1/p^d$ in general so that, for any $d$, the 2-dimensional looking graviton
propagator (\ref{gp3}) is obtained. (This is equivalent to saying that
$\eta=2-d$, or $d+\eta=2$, for arbitrary $d$.)

On the other hand, the impact of the RG effects on the diffusion process is to
replace the operator $\Delta$ by $\Delta^2$, for any $d$, since the 
cosmological constant always runs quadratically. Hence, in the fixed point
regime, eq. (\ref{trplane}) becomes
$P(T)\propto\int d^dp\,\exp\left[-p^4\,T\right]\propto T^{-d/4}$.
This $T$-dependence implies the spectral dimension
\begin{eqnarray}
\label{qegspecdims2}
{\mathcal D}_{\rm s}(d)&=&d/2
\end{eqnarray}
This value coincides with $d+\eta$ if, and only if, $d=4$. It is an intriguing
speculation that this could have something to do with the observed macroscopic
dimensionality of spacetime.

For the sake of clarity and to be as explicit as possible we described the
computation of ${\mathcal D}_{\rm s}$ within the Einstein-Hilbert truncation.
However, it is easy to see \cite{oliverfrac} that the only nontrivial 
ingredient of this computation, the scaling behavior $\Delta(k)\propto k^2$,
is in fact an exact consequence of asymptotic safety. If the fixed point 
exists, simple dimensional analysis implies $\Delta(k)\propto k^2$ at the
un-truncated level, and this in turn gives rise to (\ref{qegspecdims2}). If 
QEG is asymptotically safe, ${\mathcal D}_{\rm s}=2$ at sub-Planckian 
distances is an {\it exact} nonperturbative result for all of its spacetimes.

It is interesting to compare the result (\ref{qegspecdims}) to the
spectral dimensions which were recently obtained by Monte
Carlo simulations of the causal dynamical triangulation model of quantum
gravity \cite{ajl34}:
\begin{eqnarray}
\label{mcspecdims}
{\mathcal D}_{\rm s}(T\rightarrow\infty)&=&4.02\pm 0.1\nonumber\\
{\mathcal D}_{\rm s}(T\rightarrow 0)&=&1.80\pm 0.25
\end{eqnarray}
These figures, too, suggest that the long-distance and short-distance spectral
dimension should be 4 and 2, respectively.
The dimensional reduction from 4 to 2 dimensions is a highly nontrivial
dynamical phenomenon which seems to occur in both QEG and the discrete
triangulation model. We find it quite remarkable that the discrete
and the continuum approach lead to essentially identical conclusions in this
respect. This could be a first hint indicating that the discrete model and 
QEG in the average action formulation
describe the same physics.

\section{Summary}
\renewcommand{\theequation}{7.\arabic{equation}}
\setcounter{equation}{0}
\label{conclusio}
In the first part of this article we reviewed the asymptotic safety scenario of
quantum gravity, and the evidence supporting it coming from the average 
action approach. We explained why it is indeed rather likely that 4-dimensional
Quantum Einstein Gravity can be defined (``renormalized'') nonperturbatively
along the lines of asymptotic safety. The conclusion is that it seems
quite possible to construct a quantum field theory of the spacetime metric 
which is not only an effective, but rather a fundamental one and which is 
mathematically consistent and predictive on
the smallest possible length scales even. If so, it is not necessary to leave
the realm of quantum field theory in order to construct a satisfactory quantum
gravity. This is at variance with the basic credo of string theory, for
instance, which is also claimed to provide a
consistent gravity theory. Here a very high price has to be paid for curing
the problems of perturbative gravity, however: one has to live with infinitely
many (unobserved) matter fields.

In the second part of this review we described the spacetime structure in
nonperturbative, asymptotically safe gravity. 
The general picture of the QEG spacetimes which emerged 
is as follows. At sub-Planckian distances spacetime is a fractal of
dimensionality ${\mathcal D}_{\rm s}=4+\eta=2$. It can be thought of as a
self-similar hierarchy of superimposed Riemannian manifolds of any curvature.
As one considers larger length scales where the RG running of the
gravitational parameters comes to a halt, the ``ripples'' in the spacetime
gradually disappear and the structure of a classical 4-dimensional manifold is
recovered.

\end{document}